%% file: ASE-SRC.tex
\documentclass[sigconf]{acmart}
\usepackage{booktabs} 
\usepackage{graphicx}
\usepackage{array}
\usepackage{url}
\usepackage{fancybox}
\usepackage{multirow}
\usepackage{color}
\usepackage{colortbl}
\usepackage{fancyhdr}
\usepackage{algorithm}
\usepackage{subfig}
\usepackage{diagbox}
\usepackage{bbding}
\usepackage{placeins}
\usepackage{balance}
\usepackage{listings}

\copyrightyear{2020}
\acmYear{2020}
\setcopyright{acmcopyright}\acmConference[ASE '20]{35th IEEE/ACM International Conference on Automated Software Engineering}{September 21--25, 2020}{Virtual Event, Australia}
\acmBooktitle{35th IEEE/ACM International Conference on Automated Software Engineering (ASE '20), September 21--25, 2020, Virtual Event, Australia} \acmPrice{15.00}
\acmDOI{10.1145/3324884.3418923}
\acmISBN{978-1-4503-6768-4/20/09}

\begin{document}

	\title{Finding Ethereum Smart Contracts Security Issues \\ by Comparing History Versions} 
	
	\author{Jiachi Chen}
	\email{Jiachi.Chen@monash.edu}
	\affiliation{%
		\institution{Monash University}
	}

\begin{abstract}  
	Smart contracts are Turing-complete programs running on the blockchain. They cannot be modified, even when bugs are detected. The \emph{Selfdestruct} function is the only way to destroy a contract on the blockchain system and transfer all the Ethers on the contract balance. Thus, many developers use this function to destroy a contract and redeploy a new one when bugs are detected. In this paper, we propose a deep learning-based method to find security issues of Ethereum smart contracts by finding the updated version of a destructed contract. After finding the updated versions, we use open card sorting to find security issues. 
	
\end{abstract}


\maketitle
	
	\input{introduction}

	\input{methodology}
	\input{result}

	\input{related}

	\input{conclusion}
	\bibliographystyle{ACM-Reference-Format}
	\bibliography{ref}
	
\end{document}

%% file: introduction.tex
\section{Introduction}
\label{Introduction}

In recent years, decentralized cryptocurrencies have attracted considerable interest.  Ethereum~\cite{Ethereum_yellow_paper} is the most popular blockchain platform that supports the running of smart contracts. Smart contracts are Turing-complete programs that run on the blockchain. They cannot be modified, even when bugs are detected. 


The \emph{Selfdestruct} function~\cite{Solidity} is the only way to destroy a contract on the blockchain system and transfer all the Ethers on the contract balance. Many developers choose to add it to their smart contracts. Thus, when emergency situations happen, e.g., bugs are found by attackers, developers can use \emph{Selfdestruct} function to destruct the buggy contracts and transfer Ethers to reduce the financial loss. When bugs are patched, developers can redeploy a new contract. 

In this paper, we first crawl all the verified (open-sourced) destructed smart contracts from Etherscan~\cite{EtherScan}, a famous Ethereum smart contract explorer. Then, we download other smart contracts that are deployed by the same creators of the destructed contracts. After that, we proposed a deep learning-based method to compute the similarity between different codes. If the similarity of a contract with a destructed contract is higher than a threshold, we regard the contract is the updated version of the destructed contracts. Finally, we manually analyze the difference to summarize the security issues.

%% file: methodology.tex
\section{Methodology}

\begin{figure}
	\begin{center}
		\includegraphics[width=0.45\textwidth]{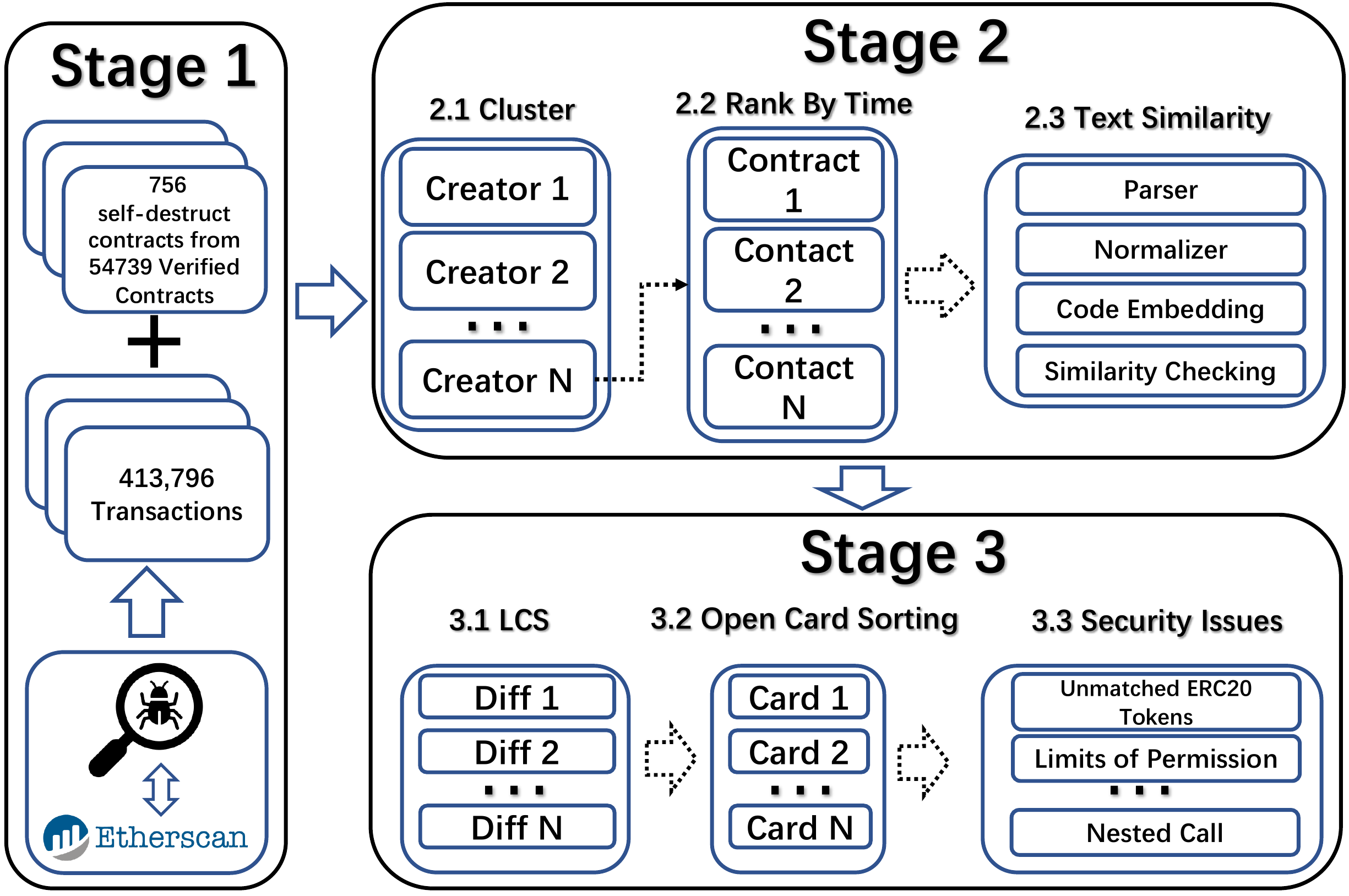} 
		\caption {Overview architecture of finding Ethereum smart contracts security issues by comparing history versions}  ~\vspace{-0.5cm}
		\label{Fig:suicdeReason}
	\end{center}
\end{figure} 

Figure~\ref{Fig:suicdeReason} depicts the detailed steps to identify security issues by comparing history versions. Our method consists of three stages. In the following parts, we introduce the details of each stage.    \vspace{-0.1cm}

\subsection{Stage 1. Data Collection}
Stage 1 is used to collect data for the following two stages. Our data contains three parts, i.e., \textit{verified contracts}, \textit{self-destruct contracts}, and \textit{contract transactions}. Verified Contracts are the open sourced smart contracts crawled from Etherscan. We totally obtained 54, 739 verified contract in our dataset. Among these smart contracts, 756 contracts are destructed smart contracts.  Transactions on Ethereum record the information of the external world interacting with the Ethereum network. In the first transaction, we can find who deployed the contract (creator), and we can find who destructed the contract (destructor) in the last transaction.  We collect all 413,796 transactions of 756 self-destruct smart contracts.

\subsection{Stage 2. Upgrade Contracts Selection}
The aim of stage 2 is to find the upgrade version of a destructed contract.  

\textbf{Step 2.1 Cluster}: We first find the creator addresses of all the 54,739 verified smart contracts through their transaction. Then, we classify the contracts into several groups according to their creator addresses. If two contracts have the same creator address, they will be classified into the same group. We only choose groups that contain self-destruct contracts. 

\textbf{Step 2.2 Rank by Time}: We first rank contracts in each group by their creation time, which can be obtained from the first transaction. Then, we can obtain several pairs; each pair is consisted of a self-destruct contract and a live contract. For example, one group contains five contracts, they are contract \textit{a,b,c,d,e} and these five contracts are ranked by creation time. Contract b and d are the self-destruct contact in these five contracts. Finally, we output four pairs, i.e., \textit{(b,c), (b,d), (b,e)} and \textit{(d,e)}. 

\textbf{Step 2.3 Text Similarity}:  We compute the code similarity between two contracts to identify whether the later created contract is the successor contract of the self-destruct contract.  We first generate ASTs (Abstract syntax trees) of each smart contract. Then, we parse the ASTs by an in-order traversal. During the parsing, all the statements of the contracts are recorded. After the parsing, we remove or replace all  the variables, punctuation marks, and different types of constants to remove the semantic-irrelevant information. Next, we embed the contracts by using Fasttext~\cite{bojanowski2017enriching} as it performs better than word2vec~\cite{rong2014word2vec}. Finally, we calculate the similarity of the contracts. If their similarity is larger than 0.6,  they might be relevant, and we assume the later created contract is the upgrade versions of the self-destruct contract. We found 436 self-destruct contracts have their upgrade contracts with 1513 \textit{\textless self-destruct contract, upgrade contract\textgreater} pairs. We note that 0.6 is a conservative threshold; we might include many irrelevant pairs in our dataset, but it will not influence our result as we conduct a manual analysis in the subsequent step. Increasing the threshold can remove some irrelevant pairs to reduce the manual effort, but it might make us miss some true matching pairs.    

\subsection{Stage 3. Security Issues Summarization}

In stages, we aim to find the security issues by comparing the difference between a self-destruct contract and its upgrade version. 

\textbf{Step 3.1 Longest Common Substring}: Longest Common Substring (LCS) algorithm is to find the longest string (or strings) that is a substring (or are substrings) of two or more strings. To reduce the manual efforts, we  use LCS to find the different parts of the two contracts. 

\textbf{Step 3.2 Open Card Sorting}: We follow the open card sorting~\cite{cardsort} approach to analyze the smart contracts and summarize the reasons why they were destructed. We create one card for each \textit{pair\textless self-destruct contract, upgrade contract \textgreater}. The detailed steps are: 

\quad \textbf{Iteration 1:} We randomly chose 20\% of the cards, and two developers with 3 years of smart contract development discussed the reason why contracts destructed. If the reason of self-destruct is unclear or irrelevant to the security issues, they omitted them from our card list. All the reasons are generated during the sorting.

\quad \textbf{Iteration 2:} The same two smart contract developers independently categorized the remaining 80\% cards into the initial classification scheme. If a new security issue is found, they discuss to verify whether the new security issue is reasonable. 

\textbf{Step 3.3 Reason Generation}:  We finally found 4 security issues, and the detailed information is shown in the following section.   

%% file: result.tex
\section{Result}
We totally find four security issues, i.e.,  Unmatched ERC20 Contract, Limits of Permission, Unchecked External Call, and Nested Call. 

\noindent {\bf 1. Unmatched ERC20 Contract.} ERC20~\cite{erc20} is the most popular standard interface for tokens in Ethereum. If the implementation of token contracts does not follow the ERC20 standard strictly, the transfer between tokens may lead to errors. For example,  ERC20 requires a transfer function to return a boolean value to identify whether the transfer is successful. Users usually use third-party tools to manipulate their tokens, and these tools capture token transfer behaviors by monitoring the standard ERC20 method~\cite{TokenScope}. If the contract does not match the ERC20 standard, the token may fail to be transferred by third-party tools. 

\noindent {\bf  2. Limits of Permission.} Since Ethereum is a permission-less network, everyone can call the function of a smart contract. However, some contracts miss checking the permission of some sensitive functions, e.g., Ether transfer, which leads to serious security issues. 

\noindent {\bf  3. Unchecked External Call.} Solidity provides a series of external call functions, e.g., address.send(), address.call(), address.delegatecall(). These methods may fail due to network errors or out-of-gas error. When errors happen, these methods will return a boolean value (\textit{False}), but never throw an exception. If callers do not check the return values of external calls, they cannot ensure whether code logic is correct.

\noindent {\bf  4. Nested Call.} Instruction CALL is very expensive (9000 gas paid for a non-zero value transfer as part of the CALL operation). If a loop body contains CALL operation but does not limit the number of times the loop is executed, the total gas cost would have a high probability of exceeding the gas limitation because the number of iterations may be high and it is hard to know its upper limit.

%% file: related.tex
\section{Related Work}
\label{related}

Chen et al.~\cite{fse-smell} define 20 smart contract defects on Ethereum by analyzing the post on StackExchange~\cite{StackExchange} and divide them into five categories, i.e., \textit{security, availability, performance, maintainability}, and \textit{reusability defects}. Oyente~\cite{oyente} is the first tool for security examination for smart contracts based on symbolic execution. Their work introduces four security issues on smart contracts, i.e., mishandled exception, transaction-ordering dependence, timestamps dependence, and re-entrancy attack.  Kalra et al.~\cite{Zeus} proposed a tool named Zeus, which can detect seven kinds of security problems; four of them are the same with Oyente; the other three issues are \textit{failed send, interger overflow/underflow}, and \textit{transaction state dependence}.

%% file: conclusion.tex
\section{Conclusion}

In this paper, we proposed a method  to identify the security issues by comparing history versions. We first crawl all verified contracts and their transactions from Etherscan. Then, we divide crawled contracts into several groups by their creators' addresses. In this case, we can find smart contracts that are created by the same authors. Next, we compute the code similarity of contracts in each group to find self-destruct contracts and their upgrade version. Finally, we summarize four security issues by using \textit{open card sorting}. 